\documentclass[final]{siamltex}
\pdfoutput=1

\usepackage{amsmath,amssymb}

\usepackage{graphicx}
\usepackage{bm}
\usepackage{braket}
\usepackage[hypertexnames=false,colorlinks=true,linkcolor=blue,citecolor=blue,
            bookmarksopen=false,bookmarks=false,
            pdfstartview=XYZ,pdffitwindow=true,pdfcenterwindow=true]{hyperref}

\newcommand*{\diff}{\mathop{}\!\mathrm{d}}
\newcommand{\e}{{\rm e}}

\newcommand{\ZSet}{\mathbb{Z}}
\newcommand{\RSet}{\mathbb{R}}

\newcommand{\wT}{\widetilde w}
\newcommand{\qat}{q_\textrm{at}}
\newcommand{\qct}{q_\textrm{ct}}
\newcommand{\qb}{q_\textrm{b}}
\newcommand{\epsi}{\varepsilon}

\newcommand{\vect}[1]{\bm{#1}}
\newcommand{\matr}[1]{\bm{#1}}

\newtheorem{rem}[theorem]{Remark}
\newtheorem{prob}[theorem]{Problem}

\newcommand{\revised}[1]{#1}
\newcommand{\rerevised}[1]{#1}

\graphicspath{{Figures/}}

\title{Snakes and Ladders in an Inhomogeneous Neural Field Model}

\author{Daniele Avitabile\thanks{Centre for Mathematical Medicine and Biology, School
of Mathematical Sciences, University of Nottingham, UK} 
\and
Helmut Schmidt\thanks{College of Engineering, Mathematics and Physical Sciences,
University of Exeter, UK}}

\begin{document}
\maketitle

\begin{abstract}
Continuous neural field models with inhomogeneous synaptic connectivities are
known to support traveling fronts as well as stable bumps of localized
activity. We analyze stationary localized structures in a neural field model with
periodic modulation of the synaptic connectivity kernel and find that they are
arranged in a \textit{snakes-and-ladders} bifurcation structure. In the case of
Heaviside firing rates, we construct analytically symmetric and asymmetric states and hence
derive closed-form expressions for the corresponding bifurcation diagrams. 
We show that the approach proposed by Beck and co-workers to analyze snaking
solutions to the Swift--Hohenberg equation remains valid for the neural field model,
even though the corresponding spatial-dynamical formulation is non-autonomous. 
We investigate how the modulation amplitude affects the bifurcation structure
and compare numerical calculations for steep sigmoidal
firing rates with analytic predictions valid in the Heaviside limit.
\end{abstract}




\section{Introduction}
\label{sec:intro}

Continuous neural field models are a common tool to investigate large-scale activity of
neuronal ensembles. Since the seminal work of Wilson and Cowan
\cite{Wilson1972q,Wilson1973o} and Amari \cite{Amari1975u,Amari1977q}, these nonlocal
models have helped understanding the emergence of spatial and spatio-temporal coherent
structures in various experimental observations. Stationary spatially-extended
patterns have been found in visual
hallucinations~\cite{Ermentrout1979y,Bressloff2001q}, while stationary localized
structures, commonly referred to as \textit{bumps} \cite{Coombes2005f}, are related to
short term (working) memory \cite{GoldmanRakic1995} and feature selectivity in the
visual cortex \cite{Ben-Yishai1995l,Hansel1997z}.
Traveling waves of neural activity are relevant for information processing
\cite{Ermentrout2001} and can be evoked \textit{in vitro} in slice preparations of
cortical \cite{Wu1999}, thalamic \cite{Kim1995} or hippocampal \cite{Miles1988}
tissue by electric stimulation (for recent reviews see \cite{Muller2012,Sato2012a}).
Furthermore, traveling waves have also been observed \textit{in vivo} in the form of
\textit{spreading depression} in neurological disorders such as migraine
\cite{Lauritzen1994}.

The simplest neural field models are (systems of) integro-differential equations
posed on the real line or on the plane. The corresponding nonlocal terms feature a
\textit{synaptic kernel}, a function that models the neural connectivity at a
macroscopic scale. For mathematical convenience, neural field models are often chosen
to be translationally invariant, that is, the synaptic kernel depends on the
Euclidean distance between points on the domain. This assumption reflects well
experiments in which cortical slices are pharmacologically prepared.
However, \textit{in vivo} experiments by Hubel and
Wiesel~\cite{Hubel1962,Hubel1968,Hubel1974,Hubel1977} revealed that a complex
microstructure is present in several areas of mammalian cortex.
In order to model this microstructure, Bressloff \cite{Bressloff2001r} incorporated a
spatially-periodic modulation of the synaptic kernel into a one-dimensional neural
field model. The translational invariance is thus broken, leading to slower
traveling waves and, for sufficiently large modulation amplitudes, to propagation
failure (a similar effect is also caused by inhomogeneities in the
input~\cite{Bressloff2003a}). In the present article we show how inhomogeneities in
the synaptic connectivity can give rise to a multitude of stable stationary bumps
which are organized in parameter space via a characteristic snaking bifurcation
structure.

The formation and bifurcation structure of stationary localized patterns has been
studied extensively in partial differential equations (PDEs) posed on
domains with
one~\cite{Woods1999c,Coullet2000u,Burke2007i,Burke2007r,Champneys2007f,Dawes2008a,Beck2009k},
two~\cite{Sakaguchi1996q,Sakaguchi1998e,Lloyd2008x,Lloyd2009b,Avitabile2010d,McCalla2010q}
and three spatial dimensions~\cite{Schneider2010w,Beaume2013a,Lo-Jacono2013a}. Most analytical
studies focus on the Swift--Hohenberg equation (or one of its variants) posed on the
real line: stationary localized solutions to the PDE connect a homogeneous
(background) state to a patterned state at the core of the domain, hence they can be
interpreted as homoclinic connections in the corresponding spatial-dynamical system.
In a suitable region of parameter space, close to the so-called \textit{Maxwell
point}, there exist infinitely many homoclinic connections, corresponding to PDE
solutions with varying spatial extent. Localized states with different
symmetries belong to intertwined solution branches that snake between two (or more)
limits and are connected by branches of asymmetric states. This bifurcation
structure was called \textit{snakes and ladders} by Burke and
Knobloch~\cite{Burke2007r}. 

It is known that snaking localized structures arise also in systems with nonlocal
terms. For instance, snaking bumps are supported by neural field models posed on
the real line~\cite{Laing2002b,Laing2003q,Coombes2003v,Faye2012j,Faye2012y} and on
the plane~\cite{Rankin2013a}, as well as the Swift--Hohenberg equation with
nonlocal terms~\cite{Morgan2014a}. In neural field models, the choice of the kernel
has an impact on the bifurcation structure~\cite{Rankin2013a}, hence it is
interesting to study how inhomogeneities affect the existence and stability of
localized states, an investigation that has been carried out very recently by
Kao et al. in the context of the Swift--Hohenberg equation~\cite{Kao2014a}.

In the present paper, we study a neural field model with a synaptic kernel
featuring a tunable harmonic inhomogeneity~\cite{Schmidt2009s,Coombes2011b}. As
pointed out by Schmidt et al.~\cite{Schmidt2009s}, the inhomogeneity gives
rise to stable bumps which do not exist in the homogeneous case. We will show here
that the synaptic modulation is also responsible for the snaking behavior of such
solutions. 

A characteristic of neural field models is that they can be conveniently
analyzed in the limit of Heaviside firing rates: for the model under consideration,
bumps can be constructed analytically, hence, following Beck et al.\cite{Beck2009k}, we
can derive closed form expressions for the snaking bifurcation curves. In addition, we
show that the Heaviside limit provides a good approximation to the case of steep
sigmoidal firing rates, for which the theory by Beck et al. can not be directly
applied.

This article is structured as follows: in Section~\ref{sec:model} we present the
neural field model and discuss stability of stationary solutions. In
Section~\ref{sec:pde} we show numerical simulations of the model in the case of
steep sigmoidal firing rates, for which an equivalent PDE formulation is available.
In Section~\ref{sec:steadyStates} we move to the
Heaviside firing rate limit, for which we discuss the construction and stability of
generic localized steady states. In Sections~\ref{sec:periodic}
and~\ref{sec:localized} we calculate explicitly periodic and localized steady states
and infer the relative bifurcation diagrams. In Section~\ref{sec:2pardiag} we discuss
how the bifurcation structure is affected by changes in the modulation amplitude.
We conclude the paper in Section \ref{sec:conclusions}.

\section{The integral model}
\label{sec:model}

We consider a neural field model of the Amari type, posed on the real line,
\begin{equation}
  \partial_t u (x,t) = -u(x,t) +\int_{-\infty}^{\infty}\, W(x,y) f(u(y,t)) \diff{y},
  \qquad 
  (x,t) \in \RSet \times \RSet^+
  \label{eq:model}
\end{equation}
where $u$ is the synaptic potential, $W$ the synaptic connectivity and $f$ a
nonlinear function for the conversion of the synaptic potential into a firing rate.
In general, both $W$ and $f$ depend upon a set of control parameters, which have
been omitted here for simplicity.

Several studies of neural field models assume translation invariance in the model
(see \cite{Bressloff2012o,Coombes2010k} and references therein), therefore the
synaptic strength $W$ depends solely on the Euclidean distance between $x$ and
$y$, that is $W(x,y)=w(|x-y|)$. A neural field of this type is said to be
\textit{homogeneous}. 

A simple way to incorporate an inhomogeneous microstructure is to
multiply the homogeneous kernel $w$ by a periodic function $A(y)$ that modulates
the synaptic connectivity and thus breaks translational invariance. 
Following Bressloff~\cite{Bressloff2001r} we choose $A(y)$ to be a simple
harmonic function and we pose
\begin{equation}
  W(x,y) = w(|x-y|) A(y), \quad \textrm{where } w(x) = \frac{1}{2} e^{x}, \quad  A(y)
  = 1+a \cos \frac{y}{\epsi}.
  \label{eq:kernel}
\end{equation}
Here, $a$ is the amplitude of the modulation and $2 \pi \epsi$ its wavelength.
With this choice, the neural field model is invariant with respect to transformations 
\begin{equation}
  x \mapsto x + 2 \pi \epsi n, \quad n \in \ZSet.
  \label{eq:symShift}
\end{equation}

In this paper we study stationary localized states of system~\eqref{eq:model}
with inhomogeneous kernels~\eqref{eq:kernel}. The firing rate $f$ will be either
a Heaviside function $f(u) =H(u-h)$, where $h$ is the firing threshold, or a
sigmoidal firing rate
\begin{equation}
  f(u) = \frac{1}{1+\exp (-\nu(u-h))}, \quad 
  \label{eq:sigmoid}
\end{equation}
with $\nu \gg 1$. In the limit $\nu \rightarrow \infty$, the sigmoidal
firing rate~\eqref{eq:sigmoid} recovers the Heaviside case. As we shall see, a
Heaviside firing rate will be more convenient for analytical calculations, whereas a
steep smooth firing rate will be employed for numerical computations.

Stationary solutions to the system~\eqref{eq:model}--\eqref{eq:kernel} satisfy
\begin{equation}
  q(x) = \int_{-\infty}^{\infty} W(x,y) f(q(y)) \diff{y}.
  \label{eq:steadyState}
\end{equation}
Linear stability is studied posing $u(x,t) = q(x) + e^{\lambda t}v(x)$, with $v
\ll 1$, and linearizing the right-hand side of
\eqref{eq:model}. This leads to the following nonlocal eigenvalue problem
\begin{equation}
  (1+\lambda) v(x) = \int_{-\infty}^{\infty}\, W(x,y) f'(q(y)) v(y) \diff{y},
  \label{eq:eigenvalueProblem}
\end{equation}
where we have formally denoted by $f'$ the derivative of $f$. This linear stability
analysis is standard in the study of localized solutions in
neural field models \cite{Coombes2005f}.

\section{PDE formulation for smooth firing rates}
\label{sec:pde}
\begin{figure}
  \centering
  \includegraphics{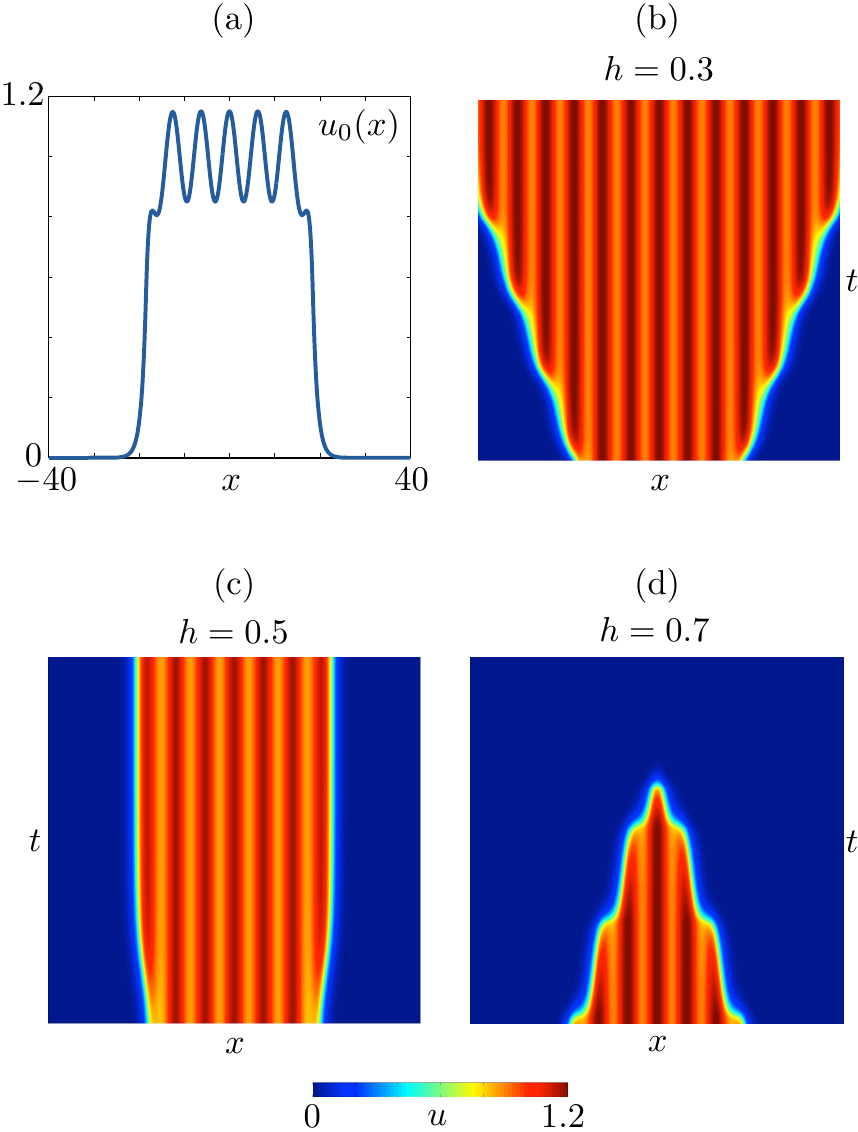}
  \caption{Time simulations of the PDE model~\eqref{eq:modelPDE} with synaptic
  kernel~\eqref{eq:kernel} and sigmoidal firing rate~\eqref{eq:sigmoid}, posed on a
  large domain $x \in [-L_x,L_x]$ with Neumann boundary conditions $\partial_x
  u(\pm L_x,t) = 0$. Panel (a): initial condition used in the simulations. Panel (b):
  the periodic core invades the domain (time runs upwards) as two pulsating fronts
  travel towards the boundary. Panel (c): a stable localized steady state is
  formed. Panel (d): the homogeneous background invades the domain. Parameters:
  $L_x=90$ (plots show an inset $(x,t) \in[-40,40] \times [0,50]$), $\nu=50$,
  $a=0.3$, $\epsilon=1$. Spatial operators are discretized via finite differences
  with $3000$ gridpoints.}
  \label{fig:timeSimulation}
\end{figure}

When $f$ is a smooth sigmoid, it is advantageous to reformulate the nonlocal
problem~\eqref{eq:model} as a local PDE, more suitable for direct numerical
simulation and numerical continuation. Following
\cite{Laing2003q,Coombes2007j,Coombes2011b,Coombes2013q,Laing2013n},
we take the Fourier transform of \eqref{eq:model}, with kernel expressed by
\eqref{eq:kernel}
\[
\partial_t \hat u(\xi,t) = - \hat u(\xi,t) + \hat w(\xi) 
\widehat{ \big( A f(u) \big)} (\xi,t), 
\]
where $\hat w(\xi) = (2\pi)^{-1}/(\xi^2+1)$. Multiplying the previous equation by
$\xi^2+1$ and taking the inverse Fourier transform we obtain
\begin{equation}
  (1 - \partial^2_x)(\partial_t u + u) = A(x) f(u),
  \label{eq:modelPDE}
\end{equation}
where the dependence of $u$ on $x$ and $t$ has been omitted for simplicity. Once
complemented with suitable initial and boundary conditions, the equation above
constitutes an equivalent PDE formulation of the model problem.
Steady states are solutions to 
\begin{equation}
  0 = (\partial^2_x - 1) q + A(x) f(q)
  \label{eq:steadyStatesPDE}
\end{equation}
and linear stability is inferred via the generalized eigenvalue problem
\begin{equation}
  (1 + \lambda) (1 - \partial^2_x) v = A(x) f'(q) v.
  \label{eq:eigenvalueProblemPDE}
\end{equation}
In passing we note that time simulations of~\eqref{eq:modelPDE} and stability
calculations~\eqref{eq:eigenvalueProblemPDE} can be carried out numerically
without forming a discretization for $(1-\partial^2_{x})^{-1}$
(see~\cite{Coombes2013q}).
In Figure~\ref{fig:timeSimulation} we show time simulations of
\eqref{eq:modelPDE} posed on the interval $[-90,90]$ with Neumann boundary
conditions, for various values of the firing rate threshold $h$. For selected
values of $h$, we find stable localized solutions, which destabilize as the
parameter is increased or decreased. Time-dependent solutions, such as the ones shown
in Figures~\ref{fig:timeSimulation}(b) and~\ref{fig:timeSimulation}(d), have been
previously analyzed by Coombes and Laing~\cite{Coombes2011b}, whereas in the present
paper we focus on the existence and bifurcation structure of stationary localized
states. 

\begin{figure}
  \centering
  \includegraphics{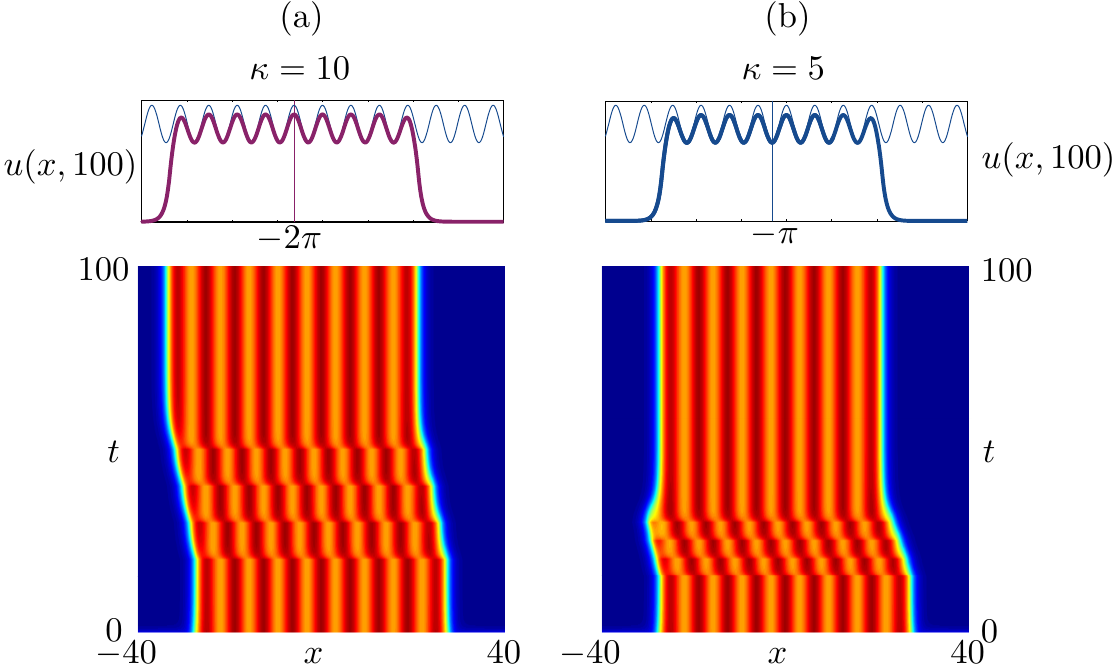}
  \caption{Time simulations of the PDE model~\eqref{eq:modelPDE} with
  instantaneous phase slips in the kernel modulation~\eqref{eq:phaseSlips}.
  Panel (a): with $\kappa=10$ we obtain a localized steady state that is
  symmetric with respect to the axis $x=-2\pi$. Panel (b): for $\kappa = 5$, the
  steady state is symmetric with respect to the axis $x=-\pi$. Parameters as in
  Figure~\ref{fig:timeSimulation}(c). In the top panels we plot reference curves that
  are spatially in phase with the inhomogeneity $A$.}
  \label{fig:timeSimulationPhaseSlips}
\end{figure}

The simulations in Figure~\ref{fig:timeSimulation} are compatible with a
\textit{snakes-and-ladders} bifurcation structure and, owing to the
spatial modulation, we expect to find stable localized states that are spatially
in-phase with $A$ and centered around its local minima and maxima.
In Figure~\ref{fig:timeSimulationPhaseSlips} we fix $h$ and perturb a localized
steady state with abrupt phase slips in the kernel modulation. More
precisely we set
\begin{equation}
A(x,t) = 1 + a \cos\bigg(\frac{x}{\epsi} + \sum_{i=1}^{4} i \frac{\pi}{2}
\chi_{[t_i,t_{i+1}]}(t) \bigg) 
\label{eq:phaseSlips}
\end{equation}
where $t_i = 10+i\kappa$ and $\chi_{[t_i,t_{i+1}]}$ is the indicator function
with support $[t_i,t_{i+1}]$. After four phase slips, we return to the original
spatial inhomogeneity $A(x) = 1 + a \cos(x/\epsi)$, which is kept constant
thereafter. Perturbations with $\kappa = 10$ elicit a localized steady state
that is symmetric with respect to the axis $x=-2\pi$, whereas shorter phase
slips, with $\kappa = 5$, give rise to states that are symmetric with respect to
the axis $x=-\pi$.

In local models supporting localized states, symmetries of the PDE are reflected
in the bifurcation structure: each snaking branch includes solutions with the
same symmetry and intertwined branches are connected by ladders of asymmetric
solutions. In one-dimensional snaking systems with spatial reversibility,
localized states can be interpreted from a spatial-dynamical systems viewpoint
and symmetries of the PDE correspond to reversers of the spatial-dynamical
system~\cite{Beck2009k,Makrides2014a}. Following this approach, we
recast~\eqref{eq:steadyStatesPDE} as a first-order non-autonomous system in $x$
\begin{equation}
\frac{\diff}{\diff x}
\begin{pmatrix}
  U_1 \\
  U_2
\end{pmatrix}
=
\begin{pmatrix}
  U_2 \\
  U_1 - A(x) f(U_1)
\end{pmatrix},
\label{eq:steadyStatesODE}
\end{equation}
where we posed $(U_1,U_2) = (q,q_x)$. Localized steady states of the nonlocal
model correspond to bounded solutions
to~\eqref{eq:steadyStatesODE} that decay exponentially as $\vert x \vert \to
\infty$. 
System~\eqref{eq:steadyStatesODE} is reversible: for each $n \in \ZSet$, we
consider the following autonomous
extension
\[
\frac{\diff}{\diff x}
\begin{pmatrix}
  U_1 \\
  U_2 \\
  U_3 
\end{pmatrix}
=
\begin{pmatrix}
  U_2 \\
  U_1 - A(U_3+n \pi \epsi) f(U_1) \\
  1
\end{pmatrix}
, \qquad n \in \ZSet
\]
with reverser
\[
\mathcal{R} \colon (U_1,U_2,U_3) \mapsto (U_1,-U_2,-U_3).
\]
\revised{
We say that a stationary state $q$ is even-symmetric (odd-symmetric) if there exists
an even (odd) integer $n$ such that $\mathcal{R}q = q$, that is, $q(x)$ is symmetric
with respect to the axis $x = n\pi\epsi$ and $n$ is even (odd). 
}
%
Conversely, we say that a solution is
asymmetric if $\mathcal{R}q \neq q$. The stationary profiles plotted in
Figures~\ref{fig:timeSimulationPhaseSlips}(a)
and~\ref{fig:timeSimulationPhaseSlips}(b) correspond to an even- and
odd-symmetric solution, respectively.
\begin{figure}
  \centering
  \includegraphics{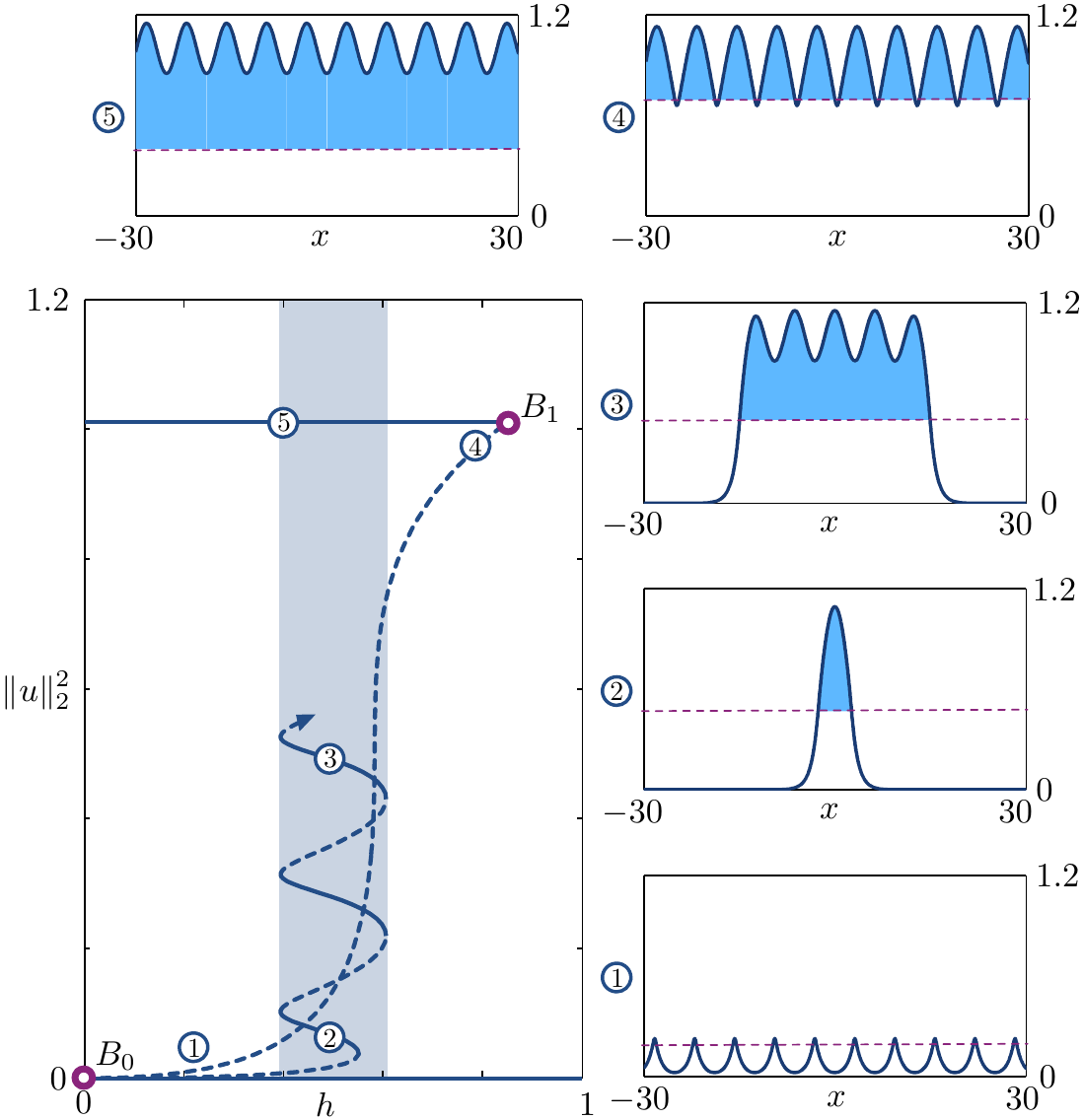}
  \caption{Branches of periodic and localized steady states of the integral model
  with inhomogeneous kernel~\eqref{eq:kernel} with $a=0.3$, $\epsi = 1$ and
  Heaviside firing rate. The bifurcation diagram is plotted in terms of the norm
  $\Vert u \Vert^2_2 = 1/(2L_x) \int_{-L_x}^{L_x} \vert u(x) \vert^2\, dx$, where
  $L_x=\pi\epsi$ for periodic solutions and $L_x=30$ for localized solutions. The
  trivial steady state $u(x)\equiv 0$ coexists with the fully periodic state for $h
  \in \big(0,1-a\epsi^2/(1+\epsi^2)\big)$. A snaking branch of even-symmetric
  localized solutions emanates from $B_0$. As we ascend the snaking diagram, more
  bumps are formed. Example solutions are plotted in the panels. For reference, we
  also plot the activity threshold $u(x) \equiv h$ (dashed magenta). There exist (not
  shown) a snaking branch of localized odd-symmetric solutions, as well as ladder
  branches connecting the two snaking branches. }
  \label{fig:heavisideSnaking}
\end{figure}

The spatial-dynamic formulation developed in~\cite{Beck2009k,Makrides2014a} for
the Swift--Hohenberg equation \rerevised{allows predictions of} snaking branches of
localized patterns from the bifurcation structure of fronts connecting the
trivial (background) state to the core state. \revised{A localized solution to
the Swift--Hohenberg PDE corrresponds to a heteroclinic orbit of the
spatial-dynamical system, \rerevised{in which the variable $x$ plays the role of
time}. If we denote by $L$ the spatial extent of the
localized solution, then the corresponding heteroclicic orbit spends a time $L$
in the proximity of the periodic core state. In the snaking bifurcation diagram
the $L_2$-norm of localized solutions, which is proportional to  $L$, is
parametrized by a control parameter of the PDE. Hence, it is possible to predict
the occurrence of snaking branches by focussing on heteroclinic orbits of the
spatial-dynamical system and studying how the time $L$ depends upon the control
parameter.}   

We can not directly apply this theory to our case, in that
system~\eqref{eq:steadyStatesODE} is non-autonomous, and $(0,0)$ is not an
equilibrium \revised{when the firing rate is sigmoidal}. However, we
shall see that the \revised{interpretation of the snaking bifurcation diagram in
terms of $L$ remains valid:}
in the limit of Heaviside firing rate, which gives rise to a non-smooth
spatial-dynamic formulation, we are able to
compute explicit expressions for connecting orbits and, hence, for the snaking
bifurcation diagram, which we partially present in
Figure~\ref{fig:heavisideSnaking}.
\revised{We construct connecting orbits directly in the integral
formulation~\eqref{eq:steadyState}, as opposed to the non-smooth, non-autonomous
spatial-dynamical formulation, as the former is more natural in the context of
neural field models.} For sigmoidal firing rates we will adopt numerical
continuation and compute snaking bifurcation branches solving the boundary-value
problem~\eqref{eq:steadyStatesPDE} and the associated stability
problem~\eqref{eq:eigenvalueProblemPDE}. 
\section{Steady states for Heaviside firing rate}
\label{sec:steadyStates}
In the case of Heaviside firing rate, localized steady states with two threshold
crossings \revised{(see solutions 2 and 3 in Figure~\ref{fig:heavisideSnaking})}
can be constructed explicitly for the inhomogeneous model and their
stability can be inferred solving a simple 2-by-2 eigenvalue problem. To each steady
state
$q$ with firing threshold $h$, we associate an active region $\mathcal{B} =
\Set{ x \in \RSet | q(x) > h}$, that is, a subset of the real line in which $q$
is above
threshold.
In the case of Heaviside firing rate, this implies that $H(q(x)) \equiv 1$ if $x \in
\mathcal{B}$ and $0$ otherwise. 
\revised{We focus on the case $\mathcal{B} = [x_1,x_2]$, for which}  Equation
\eqref{eq:steadyState} can 
be rewritten as
\begin{equation}
  q(x) = \int_{x_1}^{x_2} w(|x-y|) A(y) \diff{y}.
  \label{eq:steadyStateH}
\end{equation}
If the threshold crossings $x_{1,2}$ are known, then
\eqref{eq:steadyStateH} yields the profile of the stationary solution. The
boundaries $x_1$ and $x_2$ can be determined as functions of system parameters
by enforcing the threshold crossing conditions $q(x_1) = h$, $q(x_2) = h$.
\revised{This effectively constitutes a parametrization of $L=x_2-x_1$, as
discussed in Section~\ref{sec:pde}. As we shall see, periodic solutions with two
threshold crossings per period \revised{(such as solutions 1 and 4 in
Figure~\ref{fig:heavisideSnaking})} can also be studied with an equation similar
to~\eqref{eq:steadyStateH}.}
%

If $f$ is the Heaviside function, the nonlocal eigenvalue
problem~\eqref{eq:eigenvalueProblem} is written as
%
%
\revised{
\begin{equation}
\begin{split}
  (1+\lambda) v(x)
  & = \int_{-\infty}^{\infty} W(x,y)
  H'( q(y)) v(y) \, \diff y \\
  & = \int_{-\infty}^{\infty} W(x,y)
  \sum_{i=1}^2 \frac{\delta(y-x_i)}{\vert q'(y) \vert} v(y) \, \diff y ,
  \label{eq:eigenvalue1}
\end{split}
\end{equation}
where $\delta$ denotes the Dirac delta function.
Evaluating the integral on the righthand side of \eqref{eq:eigenvalue1}
yields 
\begin{equation}
  (1+\lambda) v(x) = \sum_{i=1}^2 \frac{A(x_i)}{\vert q'(x_i) \vert }
  v(x_i) \, w(\vert x - x_i \vert ).
  \label{eq:eigenfunction1}
\end{equation}
Setting $x=x_1,x_2$ in Equation~\eqref{eq:eigenfunction1}, we obtain the
eigenvalue problem
\begin{equation}
  (1+\lambda) \vect{\xi} = \matr{M} \vect{\xi}, \qquad 
  M_{ij} = \frac{A(x_j) w(\vert x_i - x_j \vert )}{\vert q'(x_j) \vert }, 
  \qquad i,j = 1,2.
  \label{eq:eigenvaluesQCT}
\end{equation}
for which $\{(\lambda_k,\vect{\xi}_k)\}_{k=1,2}$ can be found
explicitly. In the equation above, ${\vect \xi}_k$ has entries ${\vect \xi}_k =
(v_k(x_1),v_k(x_2))^T$, where $\{v_k(x)\}_{k=1,2}$ are eigenfunctions
satisfying~\eqref{eq:eigenfunction1}. In particular, we find
\begin{equation}
  \begin{split}
    \lambda_{1,2} =
    & -1 + \frac{1}{2} w(0) \bigg( \frac{A(x_1)}{\vert q'(x_1) \vert} +
    \frac{A(x_2)}{\vert q'(x_2) \vert} \bigg) \\
    & \pm \sqrt{ \frac{w^2(0)}{4} \bigg( \frac{A(x_1)}{\vert q'(x_1) \vert} -
    \frac{A(x_2)}{\vert q'(x_2) \vert} \bigg)^2
    + w^2(x_2-x_1) \frac{A(x_1)A(x_2)}{\vert q'(x_1) q'(x_2) \vert}
    }.
  \end{split}
  \label{eq:eigvalsExplicit}
\end{equation}
}
%
%

In the following sections we will apply this framework to both periodic
and localized solutions in the Heaviside limit. 

\begin{rem}[Number of threshold crossings]
The framework presented here can be extended to patterns with an arbitrary number of
threshold crossings;
however, throughout this paper we will restrict analytic calculations to solutions
that have only two threshold crossings, or to spatially-periodic patterns with two
threshold crossings per period. The linear stability analysis outlined here is valid
for small perturbations $v$ that have the same number of threshold crossings of $q$. 
\end{rem}
\begin{rem}[Stability of solutions with no threshold crossing] \label{rem:stab}
  Solutions that do not cross threshold are linearly stable, in that the eigenvalue
  problem~\eqref{eq:eigenvalue1} gives a single eigenvalue $\lambda = -1$.
\end{rem}

\section{Homogeneous and spatially periodic solutions for Heaviside firing rates}
\label{sec:periodic}
\begin{figure}
  \centering
  \includegraphics{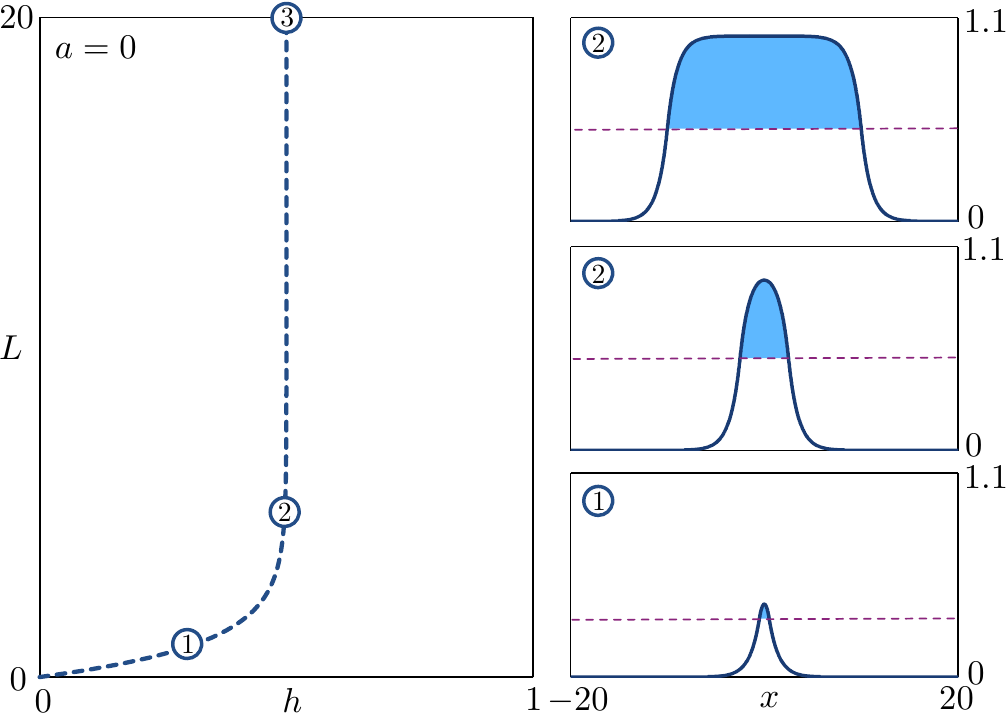}
  \caption{Bifurcation diagram for localized fronts for the homogeneous kernel. Left: branch of localized unstable 
               fronts obtained for $a=0$. We use the width $L$ of the active region as a solution measure. The branch 
               does not snake and approaches a vertical asymptote. Right: selected profiles along the branch.}
  \label{fig:noSnake}
\end{figure}

We now begin exploring steady state solutions to the integral model~\eqref{eq:model}
with inhomogeneous kernel~\eqref{eq:kernel} and Heaviside firing rate
$f(u)=H(u-h)$. 
If the kernel is homogeneous, a straightforward computation shows that
localized solutions exist and are linearly unstable. These patterns are
organized in parameter space with a non-snaking bifurcation diagram: we
integrate~\eqref{eq:steadyStateH} with $a=0$, $x_{1,2}=\pm L/2$ and obtain 
\[
q(x) = 
\begin{cases}
  1-\e^{-L/2} \cosh x  & \text{if $\vert x \vert < L/2$,} \\
  \e^{-\vert x \vert} \sinh(L/2) & \text{otherwise}
\end{cases}
\]
where $h = (1-\e^{-L})/2$. Using~\eqref{eq:eigvalsExplicit} we find
$\lambda_{1,2}\geq0$. We plot these solutions and their bifurcation diagram in
Figure \ref{fig:noSnake}.
 From now on, we will concentrate on the
more interesting case $a>0$.

Owing to the inhomogeneity, the only spatially-homogeneous solution is the
trivial state $q_0(x) \equiv 0$: posing $q(x)
\equiv \kappa$ we obtain
\[
\kappa = H(\kappa-h) \int_{-\infty}^{\infty} W(x,y) \diff y, \qquad
\]
from which we deduce $0 = \kappa < h$. The trivial solution is linearly stable for
strictly positive $h$ (see Remark~\ref{rem:stab}).

Spatially-periodic states are also supported by the integral model. In
\ref{sec:apndcell} we show that $2\pi\epsi$-periodic solutions satisfy
\begin{align}
  & q(x) = \int_{-\pi\epsi}^{\pi\epsi} \tilde w(|x-y|) A(y) f(q(y)) \diff y,
  \qquad x \in [-\pi\epsi,\pi\epsi) 
  \label{eq:2pireduced}\\
  & q(-\pi \epsi) = q(\pi \epsi),
\end{align}
where
\begin{equation}
  \widetilde{w}(x) = \frac{1}{2} \e^{-x} + \frac{\e^{-2\pi\epsi}}{1-\e^{-2\pi\epsi}} \cosh \left( x \right).
  \label{eq:modkernel}
\end{equation}
In other words, if we seek a stationary $2\pi\epsi$-periodic solution, then we
may pass from an integral equation posed on $\RSet$ to a reduced integral formulation
posed on the interval $[-\pi\epsi,\pi\epsi]$, provided that we use the amended kernel
$\wT$ instead of $w$. In passing, we note that similar conditions for periodic
solutions can be derived for generic exponential kernels.

%

We now specialize the problem~\eqref{eq:2pireduced}--\eqref{eq:modkernel} to the case
of Heaviside firing rate $f(u) = H(u-h)$, construct $2\pi\epsi$-periodic stationary
solutions and explore their bifurcation structure.
%
The simplest type of stationary periodic state of the model is the
\textit{above-threshold solution} $\qat$, that is, a solution that
lies above threshold $h$ for all $x \in \RSet$. We then formulate the following
problem: 
\begin{prob}[Above-threshold periodic solutions]
  For fixed $h,a,\epsi \in \RSet^+$, find a smooth $2\pi\epsi$-periodic function $\qat$ such that
\[
\begin{split}
   \qat(x) 
     & = \int_{-\pi\epsi}^{\pi\epsi} \wT(|x-y|) A(y) \diff y,
    \qquad x \in [-\pi\epsi,\pi\epsi), \\
   h & < \min_{x \in [-\pi \epsi, \pi \epsi)} \qat(x),
  \label{eq:periodic1}
\end{split}
\]
\end{prob}
An explicit solution $\qat$ can be computed in closed form for the specific
kernel~\eqref{eq:modkernel}, yielding
\begin{equation}
  \qat(x) = 1 + \frac{a \epsi^2}{1+\epsi^2} \cos \frac{x}{\epsi},
  \quad x \in [-\pi\epsi,\pi\epsi), 
  \label{eq:qPer}
\end{equation}
for $ h \in \big( 0,1 - a \epsi^2/(1+\epsi^2) \big)$. 
Since there are no threshold crossings, $\qat$ is stable in this interval of $h$ for
all values of $a$ and $\epsi$. In Figure~\ref{fig:heavisideSnaking}, we show an
example of $\qat$ for $a=0.3$, $\epsi = 1$ (solution label $5$).

We now turn to the more interesting case of periodic solutions that cross threshold.
The simplest of such \textit{cross-threshold} states, $\qct$, are solutions that
attain the value $h$ exactly twice in $[-\pi\epsi,\pi\epsi)$, as shown in
Figure~\ref{fig:perSolBDVarA}(a). More precisely, we derive
cross-threshold solutions as follows:
\begin{figure}
  \centering
  \includegraphics{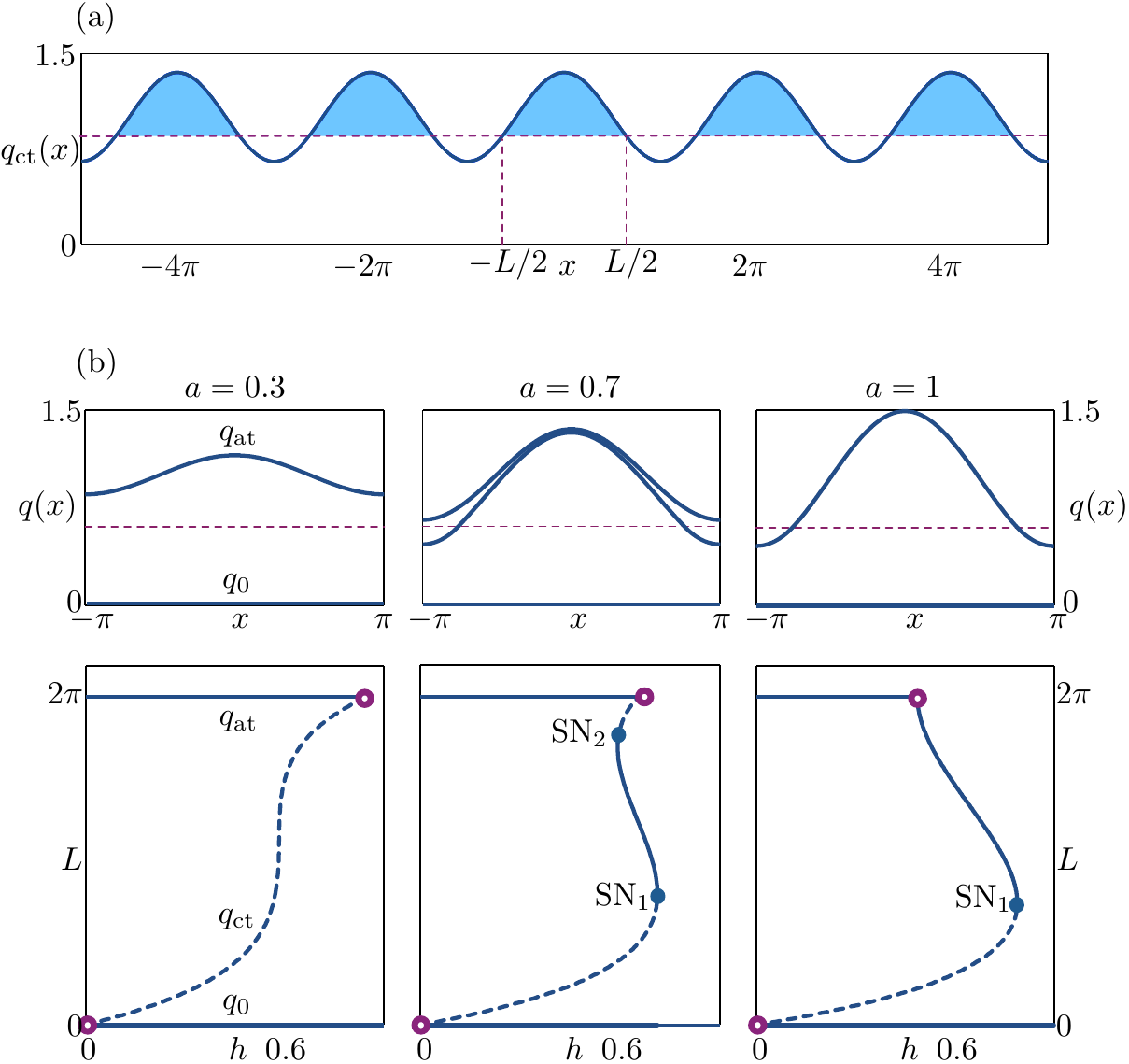}
  \caption{Periodic solutions in the Heaviside limit. (a)
  Construction of an even periodic solution $\qct$ with exactly $2$ threshold
  crossings in each period. (b) Branches of periodic solutions
  for $\epsi=1$ and various values of $a$. A branch of unstable
  solutions $\qct$ connects the branch of trivial steady states $q_0$ to the
  branch of above-threshold periodic solutions $\qat$ (bottom left); for suitable
  combinations of $a$ and $h$ the system admits three stable solutions (bottom
  center). A further increase of $a$ leads to a new bistable regime. Stable
  solutions at $h=0.6$ are shown in the top panels.}
  \label{fig:perSolBDVarA}
\end{figure}
\begin{prob}[Cross-threshold periodic solutions] \label{prob:qct}
  For fixed $h,a,\epsi \in \RSet^+$, find an even $2\pi\epsi$-periodic smooth
  function $\qct$ and a number $L \in (0,2 \pi\epsi)$ such that
  \begin{align}
    & \qct(L/2)  = h,
    \label{eq:intCond1} \\
    & \qct(x) = \int_{-L/2}^{L/2} \wT(|x-y|) A(y)\, \diff y, 
     \qquad \text{for $x \in [-\pi \epsi, \pi \epsi)$}. 
    \label{eq:2piloc}
  \end{align}
\end{prob}
The first equation implies that the threshold crossing occurs at points
$x = \pm L/2$, whereas the second one is simply derived from
Equation~\eqref{eq:2pireduced} using the identity $f(\qct(x)) \equiv 1$ for
$x \in [-L/2,L/2]$.

\begin{rem}[Bifurcation equation for periodic solutions]
  Inspecting Problem~\ref{prob:qct} we notice that the width $L$ of the
  active region of $\qct$ is a function of the threshold crossing $h$:
  combining \eqref{eq:intCond1} and \eqref{eq:2piloc} we obtain
  \begin{equation}
    h = I_\textrm{ct}(L) := 
    \int_{-L/2}^{L/2} \wT(|L/2-y|) A(y)\, \diff y.
    \label{eq:bifEqPer}
  \end{equation}
  In analogy with~\cite{Beck2009k}, we call the equation above a
  \textit{bifurcation equation} for periodic solutions
  $\qct$. Explicit formulae for the solution profile $\qct$ and the
  corresponding bifurcation equation are given in
  \ref{sec:apndexpl}.
\end{rem}


The stability of a stationary profile $(\qct,L)$ is found in a similar fashion to
what was done for stationary states in Section~\ref{sec:steadyStates}, with the
original kernel $w$ replaced by the amended kernel $\tilde w$. We find
\begin{equation}
\begin{split}
  (1+\lambda) v(x)
  & = \int_{-\pi\epsi}^{\pi\epsi} \wT(|x-y|) A(y)
  \sum_{i=1}^2 \frac{\delta(\qct(y)-x_i)}{\vert \qct'(y) \vert} v(y) \, \diff y 
  \label{eq:2pieigenvalue2}
\end{split}
\end{equation}
where $x_{1,2}=\mp L/2$. Evaluating the equation above at $x=x_{1,2}$ yields the pair of eigenvalues
\[
\lambda_{1,2} = -1 + ( \tilde w(0) \pm \tilde w(L) ) \frac{A(x_0 + L/2)}{\vert \qct'(x_0 + L/2)
\vert},
\]
where we have made use of the fact that $\vert \qct'(x) \vert$ and $A(x)$ are
even. We are now ready to study the bifurcation structure of periodic solutions
in greater detail.

\begin{figure}
  \centering
  \includegraphics{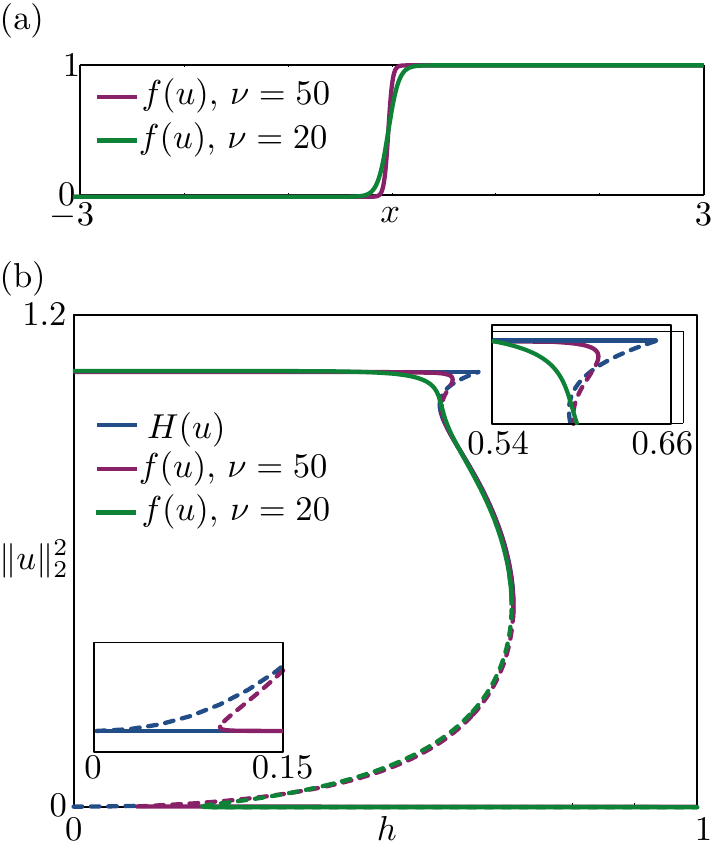}
  \caption{Branches of periodic solutions for Heaviside firing rate and steep
  sigmoidal firing rates. Panel (a): sigmoidal firing rates~\eqref{eq:sigmoid}
  with $h=0$ and $\nu=20,50$. Panel (b): bifurcation diagram of the integral
  model for $a=0.7$ and $\epsi =1$; in the Heaviside limit a branch of periodic
  solutions connects the trivial steady state $q_0$ to the above-threshold
  periodic state $\qat$; for sigmoidal firing rates the trivial steady state
  does not exist, but the branch behaves in a similar fashion, with $2$
  saddle-nodes on the branch for $\nu=20$ and $4$
  saddle-nodes on the branch for $\nu=50$.}
  \label{fig:perBDHVS}
\end{figure}

In the Heaviside limit we use
Equations~\eqref{eq:intCond1}--\eqref{eq:2pieigenvalue2} which allow us to compute
the solution profile, its activity region $L$ and its stability as a function of
$h$.
%
The resulting bifurcation diagrams are shown in Figure~\ref{fig:perSolBDVarA}(b).
The main continuation parameter is $h$ and we set $\epsi = 1$, $a\in
\{0.3,0.7,1\}$:
for small values of $a$ the trivial state $q_0$ coexists with the above
threshold solution $\qat$ for $0<h<\big( 0,1 - a \epsi^2/(1+\epsi^2) \big)$. At the grazing point
$h=1 - a \epsi^2/(1+\epsi^2)$, the above threshold solution becomes tangent to
$u(x) \equiv h$.

The branches of $q_0$ and $\qat$ are connected by a branch of
cross-threshold solutions which are initially unstable. As we increase $a$, two
saddle node bifurcations emerge on the cross-threshold branch, at a cusp, and there
exists an interval of $h$ in which $q_0$, $\qct$ and $\qat$ coexist and are
stable. As $a$ is further increased, only one saddle node persists and we have
an extended bistability region. We refer the reader to
Section~\ref{sec:2pardiag} for a more detailed study of the two-parameter
bifurcation diagram. 

We can also study the case of continuous sigmoidal firing
rates~\eqref{eq:sigmoid} using standard numerical bifurcation analysis
techniques: we find steady states $q$ solving~\eqref{eq:steadyStatesPDE} with
Neumann boundary conditions and we continue the solution in parameter space with
pseudo-arclength continuation~\cite{Govaerts2000a} using the secant code developed
in~\cite{Rankin2013a}. A comparison between bifurcation diagrams
for Heaviside and sigmoidal firing rates is presented in Figure~\ref{fig:perBDHVS}.
The solution branches are in good agreement, with the
exception of the fold points, as it can be seen in the insets.

\section{Construction and bifurcation structure of localized solutions for Heaviside
firing rates}
\label{sec:localized}

Localized steady states are solutions to \eqref{eq:model} which decay to zero as
$\vert x \vert \to \infty$ and for which the activity
region $\mathcal{B}$ is a finite disjoint union of bounded intervals
\cite{Amari1977q,Faye2012y}. 
In Figure~\ref{fig:timeSimulation} we have shown time simulations of the PDE
model~\eqref{eq:modelPDE} posed on a large finite domain with Neumann boundary
conditions and steep sigmoidal firing rate with $\nu=50$. The parameters are
chosen such that the trivial solution $q_0$ and the above-threshold periodic
solution $\qat$ are supported in the Heaviside firing rate case. As expected, stable
localized patterns are found in this region. 

In this section, we construct such patterns analytically and study their
stability. As it was done in Section~\ref{sec:periodic}, we will perform analytical
or semi-analytical calculations
 in the Heaviside limit,
whereas we will employ numerical continuation for sigmoidal firing rates.

As seen in Section~\ref{sec:steadyStates}, a generic bump $\qb$ with active
region $\mathcal{B} = (x_1,x_2) \subset \RSet$ satisfies, in the Heaviside
limit,
\begin{equation}
  \qb(x) = \int_{x_1}^{x_2} w(|x-y|) A(y) \diff y. 
  \label{eq:locstat}
\end{equation}
Without loss of generality, we pose $x_{1,2} = x_0
\mp L/2$. We note that if $L=0$ then $\qb$ coincides 
with the trivial solution. In analogy with the periodic 
case, we find a localized solution as follows:
\begin{prob}[Localized solutions]\label{prob:loc}
  For fixed $h,a,\epsilon \in \RSet^+$ find a smooth function
  $\qb$ and scalars $x_0 \in \RSet$, $L \in \RSet^+$, such that
  \begin{align}
    & \qb(x_0 - L/2)  = h, \label{eq:locThreshM} \\
    & \qb(x_0 + L/2)  = h, \label{eq:locThreshP} \\
    & \qb(x) = \int_{x_0-L/2}^{x_0+L/2} w(|x-y|) A(y)\, \diff y, 
    \qquad x \in \RSet.  \label{eq:locSol}
  \end{align}
\end{prob}
\begin{rem} In the problem above we do not enforce explicitly asymptotic
  conditions for $\qb$, since they are implied by~\eqref{eq:locSol} for our particular
  choice of $w$ and~$A$. Indeed, let $A_\ast = \max_{x \in \RSet} \vert
  A(x) \vert $, then
  \[
    0 \le \vert \qb(x) \vert \le A_\ast \int_{x_0-L/2}^{x_0+L/2}
    w(|x-y|) \, \diff y,
  \]
  hence $\vert \qb(x) \vert \to 0$ as $\vert x \vert \to \infty$.
\end{rem}

\begin{figure}
  \centering
  \includegraphics{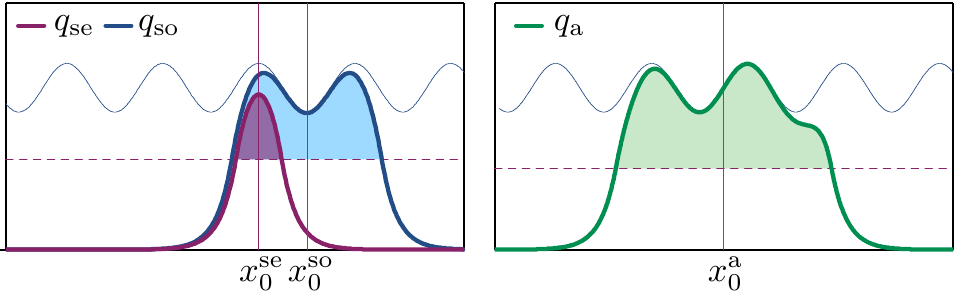}
  \caption{Examples of localized stationary solutions constructed in the Heaviside limit
  (dashed magenta line indicates the firing threshold $h$).
  Left: symmetric solutions satisfy $q(x-x_0) = q(x_0-x)$ 
  where $x_0 = n\pi \epsi$, $n \in \ZSet$. Even- (Odd-) symmetric solutions,
  $q_\textrm{se}$ ($q_\textrm{so}$), are characterized by $n$ even (odd). The
  profiles in the pictures are constructed for $h=0.55$, $a=0.3$, $\epsi=1$. As expected,
  peaks of localized solutions are in phase with peaks of the periodic solution
  $\qat$. Right: localized asymmetric solutions satisfy
  $\Psi_\textrm{asym}(L;\epsi) =0$ (see Equation~\eqref{eq:psiAsym}); for these
  patterns $x_0$ varies in a continuous interval. The asymmetric solution
  is constructed for $h=0.5$, $a=0.3$, $\epsi=1$.}
  \label{fig:exLocSols}
\end{figure}
In Section~\ref{sec:pde} we discussed symmetric and asymmetric solutions in the
context of spatial-dynamical systems of the PDE associated with the integral model.
Equivalently, a solution is symmetric if $\qb(x-x_0) = \qb(x_0-x)$ and asymmetric otherwise.
Problem~\ref{prob:loc} does not provide a direct way to distinguish between symmetric
and asymmetric states, but it can be reformulated so as to avoid this limitation.
Each solution $(\qb,x_0,L)$ to Problem~\ref{prob:loc} is such that $\qb(x_0-L/2) =
\qb(x_0+L/2)$, which can be written as
\begin{align}
  &  \Psi_\textrm{sym}(x_0) \Psi_\textrm{asym}(L)= 0, 
    \label{eq:symCond} \\
  \intertext{where}
  & \Psi_\textrm{sym}(x_0) = \sin \bigg( \frac{x_0}{\epsi} \bigg),
    \label{eq:psiSym} \\
  & \Psi_\textrm{asym}(L) = (1-\e^{-L}) 
   \cos \bigg( \frac{L}{2\epsi} \bigg) 
   - (1+\e^{-L}) \epsi \sin \bigg( \frac{L}{2\epsi} \bigg).
    \label{eq:psiAsym} 
\end{align}
Crucially, \eqref{eq:symCond} holds if either $\Psi_\textrm{sym} = 0$ or $
\Psi_\textrm{asym} = 0$,
so we are now ready to construct symmetric and asymmetric localized solutions as
follows:
\begin{prob}[Symmetric and asymmetric localized solutions]\label{prob:symLoc}
  For fixed $h,a,\epsilon \in \RSet^+$, find a smooth nonnegative function $\qb$ and
  scalars $x_0 \in \RSet$, $L \in \RSet^+$,
  such that
  \begin{align}
    & \Psi_\textrm{sym}(x_0) = 0, \qquad \text{(or $\Psi_\textrm{asym}(L)=0$)} \label{eq:SymAsymCond} \\
    & \qb(x_0 + L/2)  = h, \label{eq:locThreshP2} \\
    & \qb(x) = \int_{x_0-L/2}^{x_0+L/2} w(|x-y|) A(y)\, \diff y, 
    \qquad x \in \RSet.  \label{eq:locSol2}
  \end{align}
\end{prob}

In symmetric states, the symmetry condition~\eqref{eq:psiSym} fixes the value of
$x_0$; more precisely we have $x_0 = n \pi \epsi$ for $n \in \ZSet$, therefore we
distinguish between even- and odd-symmetric solutions, depending on the value of
$n$. On the other hand, in asymmetric states the width $L$ is fixed by the
asymmetry condition~\eqref{eq:psiAsym} and $x_0$ is not restricted
to assume discrete values.

For our choice of the connectivity function $w$ and modulation $A$ we derive
closed-form expressions for symmetric and asymmetric localized states.

For the profile of symmetric solutions we find
\begin{equation}
q_\textrm{b}(x) = 
  \begin{cases}
  \displaystyle{1 + \frac{a\epsilon^2}{1+\epsilon^2} \cos\frac{x}{\epsilon} -  \Theta_1(L;x_0) \cosh\left(x_0-x\right) }  &  \text{if  $|x-x_0|<L/2$,} \\
  \displaystyle{\Theta_2(L;x_0) \,\exp{(-|x-x_0|+L/2)} } &
  \text{otherwise,}
  \end{cases}
  \label{eq:explicitBumpSym}
\end{equation}
where the auxiliary functions $\Theta_1$ and $\Theta_2$ are given by
\begin{equation*}
  \Theta_1(L;x_0) = \left[ 1+ \frac{a\epsilon}{\sqrt{1+\epsilon^2}} \cos\frac{x_0}{\epsilon} 
  \cos\left( \frac{L}{2\epsilon}+\Phi\right) \right] \e^{-L/2} ,
\end{equation*}
\begin{equation*}
  \begin{split}
  \Theta_2(L;x_0) = & \frac{1-\e^{-L}}{2} +\frac{a}{2} \frac{\epsi}{\sqrt{\epsi^2 +1}} \cos \frac{x_0}{\epsi} \\ 
  &\times \left[ \cos\left(\frac{L}{2\epsi} - \Phi \right) - \e^{-L} \cos\left(\frac{L}{2\epsi} +\Phi \right) \right].
  \end{split}
\end{equation*}
In the above expressions we posed $\Phi = \arctan \epsilon^{-1}$ and we
exploited the fact that $\sin(x_0/\epsilon) = 0$. 

Similarly, for asymmetric solutions
we obtain
\[
q_\textrm{b}(x) = 
  \begin{cases}
  \displaystyle{1 + \frac{a\epsilon^2}{1+ \epsilon^2} \cos\frac{x}{\epsilon} -
  \Lambda_1(x,x_0;L)} 
  & \text{if $|x-x_0|<L/2$,} \\
  \displaystyle{  \Lambda_2(x_0;L)\, \exp(-|x-x_0|+L/2) } 
  & \text{otherwise,}
  \end{cases}
  \label{eq:explicitBumpAsym}
\]
with auxiliary functions $\Lambda_1$ and $\Lambda_2$ given by
\begin{equation*}
  \begin{split}
  \Lambda_1(x,x_0;L) = \frac{a\epsilon}{\sqrt{1+\epsilon^2}} \e^{-L/2} 
  \bigg\{ \sinh(x_0-x) \sin\frac{x}{\epsilon}\sin\left(\frac{L}{2\epsilon}+\Phi\right) \\ 
  + \cosh(x_0-x) \cos\frac{x}{\epsilon}\cos\left(\frac{L}{2\epsilon}+\Phi\right)  \bigg\},
  \end{split}
\end{equation*}
\begin{equation*}
  \Lambda_2(x_0;L) = \frac{1-\e^{-L}}{2} \bigg[ 1 + a \cos \bigg( \frac{x_0}{\epsi} \bigg) \cos
      \bigg( \frac{L}{2\epsi} \bigg) \bigg].
\end{equation*}

Examples of symmetric and asymmetric localized solutions
are plotted in Figure~\ref{fig:exLocSols}. These patterns are computed in a
region of parameter space where the trivial solution $q_0$ and the periodic
above-threshold $\qat$ solution coexist. 
As expected, localized solutions are in-phase with the inhomogeneity $A$.

\begin{rem}[Bifurcation equation for localized solutions]
  Similarly to the periodic case, $h$ is related to $L$ and $x_0$ via a bifurcation
  equation.
  For a solution $(\qb,x_0,L)$ of Problem~\ref{prob:symLoc},
  we find the general expression
  \[
    h = I_\textrm{b}(L,x_0) := 
    \int_{x_0-L/2}^{x_0+L/2} w(|x_0+L/2-y|) A(y)\, \diff y.
  \]
  which can be specialized for the symmetric and asymmetric cases as follows:
  \begin{align}
      h = I_\textrm{sym}(L;x_0) := & \Theta_2(L;x_0) 
    \label{eq:bifEqSymLoc} \\
      h = I_\textrm{asym}(x_0;L) := & \Lambda_2(x_0;L),
    \label{eq:bifEqAsymLoc} 
  \end{align} 
  where $\Theta_2$ and $\Lambda_2$ are auxiliary functions defined above. In the
  bifurcation function $I_\textrm{sym}$
  the value of $x_0$ is fixed by the
  condition $\Psi_\textrm{sym}(x_0) = 0$, hence $\cos(x_0/\epsilon)=\pm 1$.
  Similarly, $L$ is fixed in the expression of $I_\textrm{asym}$ and its value
  is determined by $\Psi_\textrm{asym}(L) = 0$.
\end{rem}
\begin{figure}
  \centering
  \includegraphics{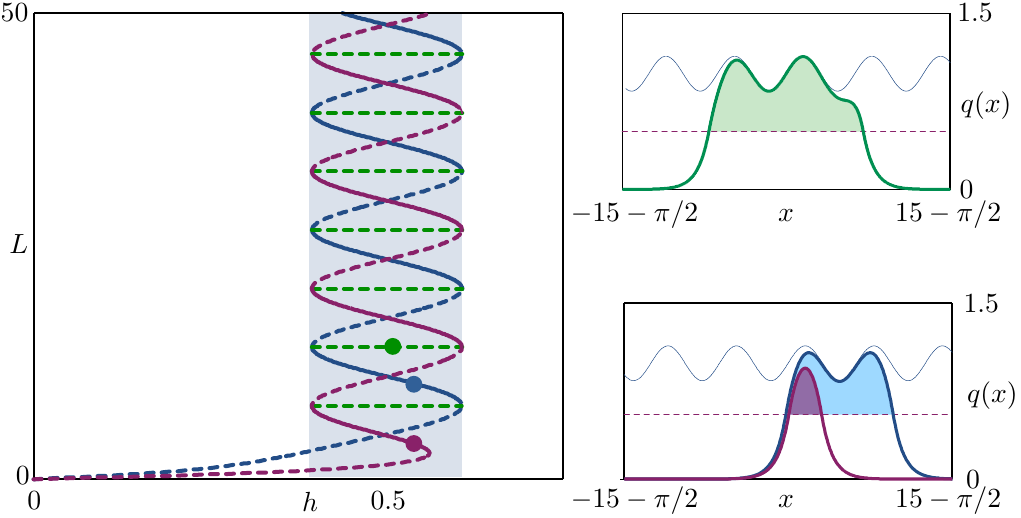}
  \caption{Snakes and ladders computed in the Heaviside limit.  Selected
  profiles along the branch are reported on the right (they correspond to
  the ones in Figure~\ref{fig:exLocSols}). Parameters as in Figure~\ref{fig:exLocSols}.}
  \label{fig:snakesAndLadderHeaviside}
\end{figure}

Following~\cite{Beck2009k}, we notice that the bifurcation
equation~\eqref{eq:bifEqSymLoc} is a parametrization of snaking branches of
even- and odd-symmetric solutions, whereas equation~\eqref{eq:bifEqAsymLoc} is a
parametrization of ladder branches of asymmetric solutions: indeed both $x_0$ and
$L$ depend on $h$, as they solve Problem~\ref{prob:symLoc}. In this case,
however, the bifurcation equations are available in closed form so we can proceed
directly to plot snakes and ladders. In Figure~\ref{fig:snakesAndLadderHeaviside}, we
fix $a$ and $\epsi$, construct localized solutions and plot their bifurcation diagrams
as loci of points on the $(L,h)$-plane that satisfy the bifurcation
equations. In particular we use $I_\textrm{sym}(L;0)$ and
$I_\textrm{sym}(L;\pi\epsi)$
to plot representative branches of even- and odd-symmetric solutions, respectively.
As expected, in the limit for large $L$, $I_\textrm{sym}$ is well approximated by a
cosinusoidal function. On the other hand, ladders are found using
$I_\textrm{asym}(x_0;L)$, where $L$ satisfies the asymmetry
condition $\Psi_\textrm{asym}(L) = 0$.

\begin{figure}
  \centering
  \includegraphics{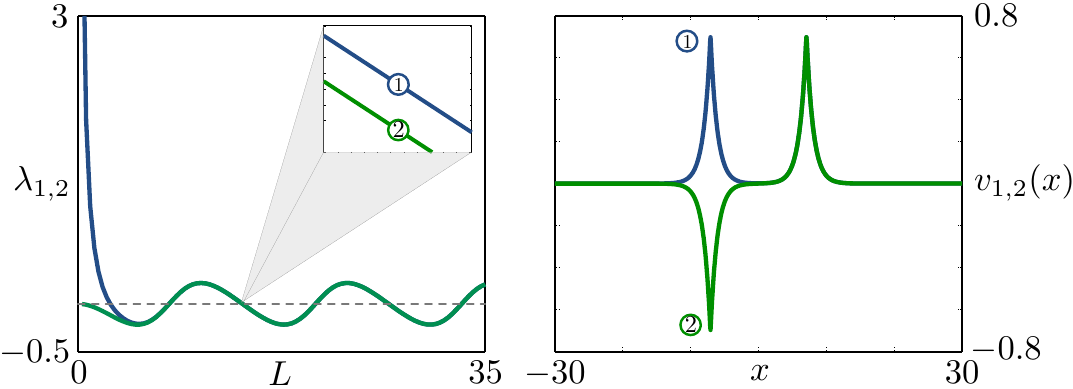}
  \caption{Eigenvalues and eigenfunctions of localized even-symmetic solutions. Left:
  eigenvalues of localized even-symmetric solutions along the snaking branch of
  Figure~\ref{fig:snakesAndLadderHeaviside}
   (see
  Equation~\eqref{eq:eigvalSym}). Right: eigenfunctions of selected states, in the
  proximity of a saddle node and a pitchfork on the snaking branch.}
  \label{fig:eigenvalues}
\end{figure}

The stability problem of a localized state $(\qb,x_0,L)$ is determined following
the scheme outlined in Section~\ref{sec:model}: we use
\revised{Equation~\eqref{eq:eigvalsExplicit}}, with threshold
crossings $x_{1,2}=x_0 \mp L/2$. 
For symmetric solutions we find
\begin{equation}
\lambda_{1,2} = -1 + ( w(0) \pm w(L) ) \frac{A(x_0 + L/2)}{\vert \qb'(x_0 + L/2) \label{eq:eigvalSym}
\vert}.
\end{equation}
In Figure~\ref{fig:eigenvalues} we plot eigenvalues $\lambda_{1,2}$,
along the even-symmetric snaking branch for $n=0$. The results show that solutions on
this branch undergo a sequence of saddle-nodes and pitchfork bifurcations, as
indicated by the corresponding eigenfunctions. Similar results (not shown) are found
for odd-symmetric states.

For asymmetric solutions we obtain
\begin{equation}
  \lambda_{1,2} = \frac{\e^{-L}}{1-\e^{-L}} \pm \frac{\Gamma(L)}{2|\qb'(x_0 + L/2)|},
  \label{eq:eigvalAsym}
\end{equation}
where
\begin{equation}
  \Gamma(L) = \sqrt{ \left( 1+\e^{-2L} \right) \left( a \sin\frac{x_0}{\epsi}
  \sin \frac{L}{2\epsi} \right)^2 + \e^{-2L} \left(1+a\cos \frac{x_0}{\epsi}
  \cos \frac{L}{2\epsi} \right)^2 }.
\end{equation}
Here we have made use of the fact that, with our choice of the synaptic kernel, we
have
\begin{equation}
  |\qb'(x_0-L/2)| = |\qb'(x_0+L/2)| = I_\textrm{asym}(x_0;L),
  \label{eq:equalityQprimeQ}
\end{equation}
which is found by differentiating \eqref{eq:explicitBumpAsym}.
Further, we note that 
\begin{equation}
  \Gamma(L) \geq \e^{-L} \left(1+a\cos \frac{x_0}{\epsi}
  \cos \frac{L}{2\epsi} \right).
  \label{eq:inequalityGammaL}
\end{equation}
By using \eqref{eq:bifEqAsymLoc}, \eqref{eq:equalityQprimeQ} and \eqref{eq:inequalityGammaL} we see
that the eigenvalues $\lambda_{1,2}$ are such that $\lambda_1 > 0$
and $\lambda_2 \leq 0$. As a consequence, all asymmetric solutions are linearly
unstable. For completeness, we find values of $h$ at which pitchfork
bifurcations are attained: such points can also be computed analytically by
setting $I_\textrm{sym}=I_\textrm{asym}$ and obtaining
\[
  h = \frac{1-\e^{-L}}{2}
  \left(1+a\cos\frac{x_0}{\epsilon}\cos\frac{L}{2\epsilon} \right), \quad
  \cos(x_0/\epsilon) = \pm 1
\]
at which
\begin{equation}
  \lambda_1 = \frac{2\e^{-L}}{1-\e^{-L}},\qquad \lambda_2 = 0.
\end{equation} 

The snake-and-ladder bifurcation structure derived here for Heaviside firing rates is
also found in the case of steep sigmoidal firing rates: in particular, we have
performed numerical continuation for the firing rate function~\eqref{eq:sigmoid} with
$\nu=50$ and found an analogous bifurcation diagram (not shown).

\section{Changes in the modulation amplitude}
\label{sec:2pardiag}
\begin{figure}
  \centering
  \includegraphics{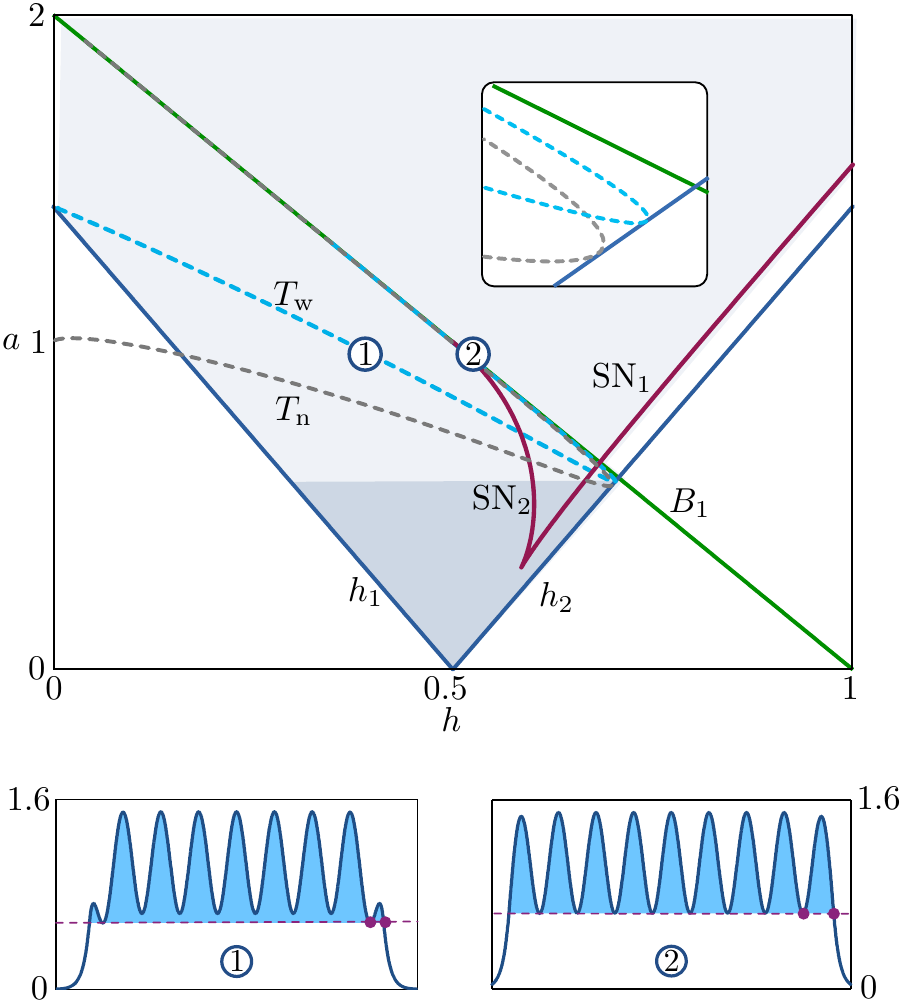}
  \caption{Two-parameter bifurcation diagram for Heaviside firing rates in the
  $(a,\epsi)$-plane. Top: We plot the grazing curve $B_1$ (in green), below which we have
  bistability between $q_0$ and $\qat$, curves of snaking limits $h_{1,2}$ (blue),
  curves of saddle nodes bifurcations of $\qct$, $\textrm{SN}_{1,2}$ (magenta) and
  lines at which a simple threshold crossing is followed by a tangency in wide
  ($T_\textrm{w}$, dashed blue line, \revised{see also patterns 1 and 2}) and narrow ($T_\textrm{n}$, dashed grey
  line) solutions. We find a simple snake and ladder scenario in the shaded blue
  area and a more complicated snaking scenario in the shaded grey area. Bottom:
  selected solution profiles on the curve $T_\textrm{w}$.}
  \label{fig:2ParBD}
\end{figure}
The framework developed in the previous Sections can be employed to study
two-parameter bifurcation diagrams. So far, we have fixed the parameters
$a$, $\epsi$ and used $h$ as our main continuation parameter. It is interesting to
explore how variations in secondary parameters affect the snaking branches.
In~\cite{Kao2014a}, the authors explore variations in the spatial scale of the
heterogeneity for the Swift--Hohenberg equation. Here, we concentrate on the
amplitude $a$ of the heterogeneity $A(x)$ for the integral neural field model.
Following the previous sections, we 
study the Heaviside case analytically and then present numerical simulations for
the steep sigmoid case.

We begin by considering Heaviside firing rate and outlining the region of
parameter space where the trivial steady state $q_0$ and the above-threshold
periodic solution $\qat$ coexist and are stable, that is, we follow the grazing
point $B_1$ in Figure~\ref{fig:heavisideSnaking} in the
$(a,h)$ plane. The curve is found by imposing the tangency condition $h=\min_{x
\in [-\pi\epsi,\pi\epsi)} \qat(x)$, which combined with Equation~\ref{eq:qPer}
gives the locus of points
\begin{equation}
  a = (1-h)\frac{1+\epsi^2}{\epsi^2}, \qquad h \in (0,1),
  \label{eq:grazing}
\end{equation}
\begin{figure}
  \centering
  \includegraphics{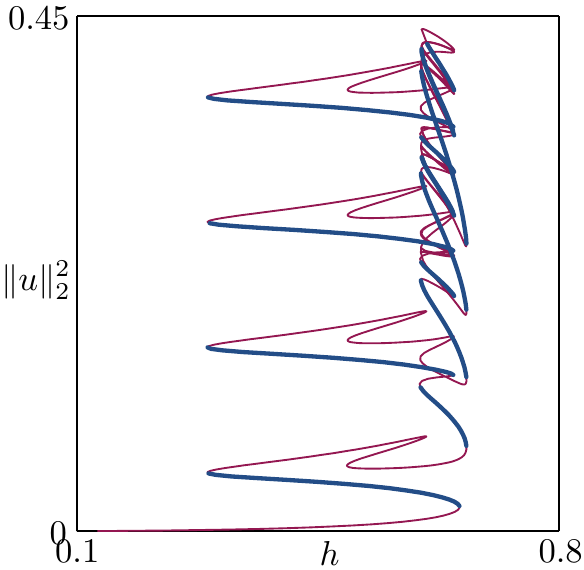}
  \caption{Branch of even solutions for sigmoidal firing rate with $a=0.6$, $\nu=50$,
  $\varepsilon=1$. Stable (Unstable) branches are indicated with
  thick blue (thin magenta) lines.}
  \label{fig:snake_a_06}
\end{figure}

In Figure~\ref{fig:2ParBD} we present a two-parameter bifurcation diagram and
indicate with a green line the locus of grazing points~\eqref{eq:grazing}:
$\qat$ and $q_0$ coexist and are stable if $(a,h)$ is
below the green line.
Next, we compute the snaking limits, for large $L$, as functions of $h$ and $a$.
We use the bifurcation equation for symmetric localized states,
Equation~\eqref{eq:bifEqSymLoc}, and find in the limit for large $L$ the
following snaking limits
\[
  h_{1,2} = \frac{1}{2} \bigg( 1 \pm \frac{a \epsi}{\sqrt{1+\epsi^2}} \bigg)
\]
These curves are plotted in Figure~\ref{fig:2ParBD} (solid blue lines). 
Further, we compute
 the loci of saddle-node
bifurcations of the cross-threshold solutions $\qct$ (which are labeled $SN_1$ and
$SN_2$ in Figure~\ref{fig:perSolBDVarA}) by solving for $(a,h)$ the following
system
\begin{align*}
  & h - I_\textrm{ct}(L; x_0) = 0 \\
  & \frac{\diff}{\diff L} I_\textrm{ct}(L; x_0) = 0
\end{align*}
The loci of saddle-node bifurcations are plotted with dark magenta lines in
Figure~\ref{fig:2ParBD}. The area between these two curves identifies a region
in which $\qct$, $\qat$ and $q_0$ coexist and are stable. In passing we note
that the curve for $SN_2$ intersects the curve for the grazing point $B_1$ at
$a=1$.

\begin{figure}
  \centering
  \includegraphics{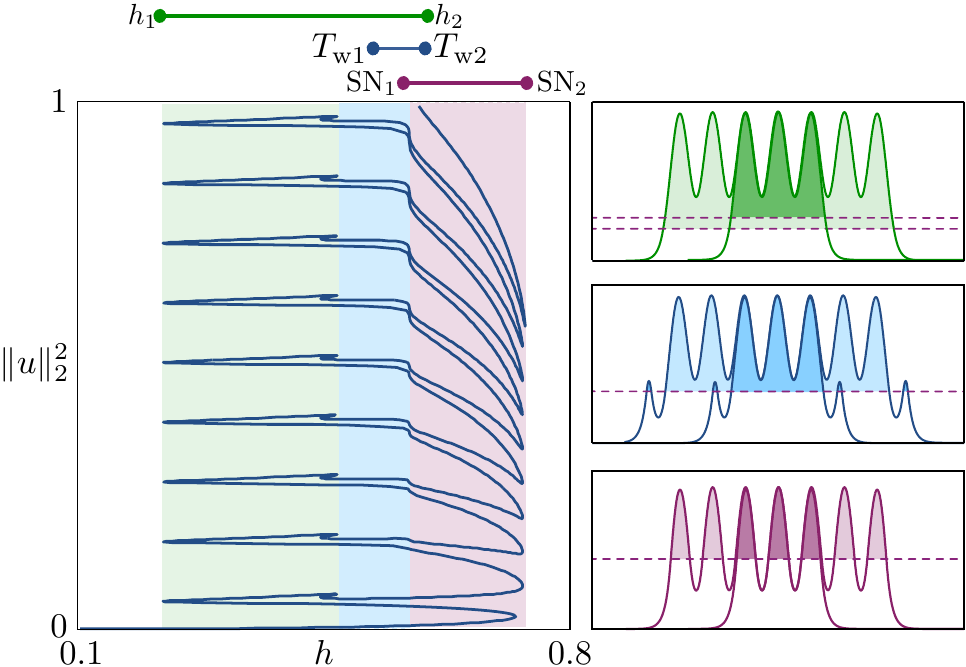}
  \caption{Bifurcation diagram for $a=0.8$ $\nu=50$, $\varepsilon=1$.
  The snaking branch is composed by solutions with two (green), six (blue) or more
  (magenta) threshold crossings. The snaking structure reflects these three types of
  solutions and their occurrence is predicted adequately by the two-parameter
  bifurcation analysis for Heaviside case (reference intervals are reported on top of
  the bifurcation diagram). Stability is not indicated and a second intertwined branch
  of odd localized states is also found (not shown).}
  \label{fig:snake_a_08}
\end{figure}

We found a snake-and-ladder bifurcation structure, as discussed in
Section~\ref{sec:localized}, in a wedge delimited by the lines $h_1$ and $h_2$
for $a \lesssim 0.57$ (dark blue area in Figure~\ref{fig:2ParBD}). Snaking
branches in this region are formed of solutions with exactly two threshold crossings
at $x_0 \mp L/2$. However, there exist snaking branches of solutions with
more threshold crossings. An example is given for the steep sigmoidal case for
$a = 0.6$ in \revised{Figure~\ref{fig:snake_a_06}}: the snaking branch collides with
neighbouring branches of solutions with multiple crossings and give rise to an
intricate bifurcation structure.

\begin{figure}
  \centering
  \includegraphics{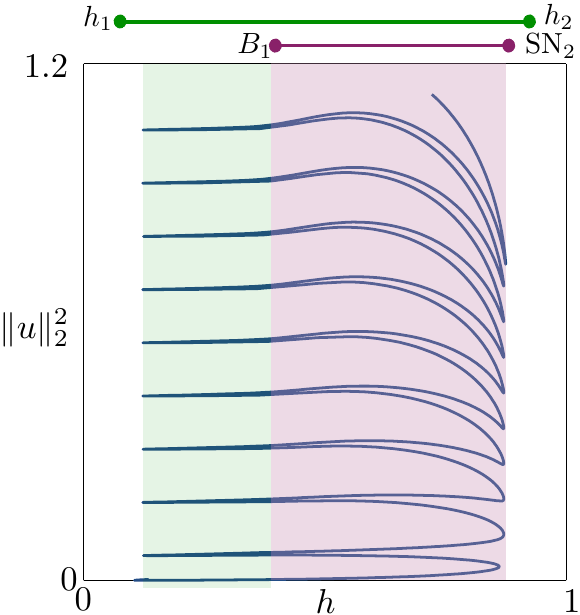}
  \caption{Bifurcation diagram for $a=1.2$, $\nu=50$, $\varepsilon=1$. Reference
  intervals for the Heaviside case are reported on top of the bifurcation diagram.
  The branches occupy a wide region of parameter space, as we expect from the
  extended bistability region of periodic solutions for high values of $a$ (see panel
  (b) for $a=1$ in Figure~\ref{fig:perSolBDVarA})}
  \label{fig:snake_a_1p2}
\end{figure}

In order to understand the occurrence of such curves we return to the Heaviside case
and concentrate on the even- and odd-symmetric solutions featuring a threshold
crossing followed by a threshold tangency at a local minimum (for an example with
large $L$, see pattern $1$ in Figure~\ref{fig:2ParBD}). More precisely, we denote by
$x_\ast$ the point with largest absolute value at which $\qb$ attains a local minimum
and solve for $(a,h,x_\ast)$ the system
\begin{align}
  & \qb(x_\ast) - q_\textrm{b}(x_0 + L/2) = 0 \label{eq:tan1} \\ 
  & \qb(x_\ast) - h                       = 0 \label{eq:tan2} \\
  & \qb'(x_\ast)                          = 0 \label{eq:tan3} 
\end{align}
where $q_\textrm{b}$ is given by Equation~\eqref{eq:locSol2}. We follow solutions
to the system above as $L$ varies in a given range and show the corresponding loci of
solutions in the $(a,h)$-plane in Figure~\ref{fig:2ParBD}: the dashed curve
$T_\textrm{w}$ contains solutions to \eqref{eq:tan1}--\eqref{eq:tan3}
with a wide active domain ($L$ varies
approximately between $45$ and $57$), whereas $T_\textrm{n}$ corresponds to solutions
with a narrow active domain ($L$ varies approximately between $0.7$ and $12$). Even
though $T_\textrm{w}$ and $T_\textrm{n}$ are not loci of bifurcations, they are
indicative of regions of parameter space where solutions with multiple threshold
crossings may occur.

In Figure~\ref{fig:snake_a_08} we show a bifurcation diagram for
$a = 0.8$ for the steep sigmoid: the snaking branch is composed by solutions with two
(green), six (blue) or more (magenta) threshold crossings. The snaking structure
reflects these three types of solutions and their occurrence is predicted adequately
by the two-parameter bifurcation analysis for Heaviside case (reference intervals are
reported on top of the bifurcation diagram of Figure~\ref{fig:snake_a_08}). Stable
and unstable branches alternate in the usual manner and an intertwined branch of
localized odd solutions exists as well (not shown). A similar scenario, with an
even wider snaking diagram, is found for $a=1.2$ (see
Figure~\ref{fig:snake_a_1p2}): for large modulation amplitudes, the bifurcation
diagram also contains cross-threshold solutions, but this time their occurrence
is marked by the grazing point $B_1$ and the saddle nodes $SN_2$ (see also the
bifurcation diagram for $a=1$ in Figure~\ref{fig:perSolBDVarA}(b)).
 
\section{Conclusions}
\label{sec:conclusions}

In the present paper we have studied the existence and bifurcation structure of
stationary localized solutions to a neural field model with inhomogeneous synaptic
kernel. For Heaviside firing rates, we computed localized as well as 
spatially-periodic solutions and we followed them in parameter space.
We recovered the classical snakes and ladders structure that
is found in the one-dimensional Swift--Hohenberg equation as well as previous
studies in neural field models: for our model, however, 
both solutions and bifurcation equations are found analytically. Since
linear stability can also be inferred with a simple calculation, it is possible to
draw the snaking bifurcation diagrams analytically or semi-analytically
(using elementary quadrature rules for the integrals).

Interestingly, we found that the interpretation of the snake and ladder structure
proposed by Beck and co-workers~\cite{Beck2009k} and extended by Makrides and
Sandstede~\cite{Makrides2014a} is valid for the specific inhomogeneous case
presented here, for both Heaviside and sigmoidal firing rates: it seems plausible
that their framework could be extended to tackle
the corresponding non-autonomous spatial-dynamical
formulation~\eqref{eq:steadyStatesODE}.

With reference to the particular system presented here, we found that a harmonic
modulation with an $\mathcal{O}(1)$ spatial wavelength promotes the formation of
snaking localized bumps and we note that these structures are driven entirely by
the inhomogeneity: in the translation-invariant case, $a=0$, the
system supports localized fronts belonging to a non-snaking branch (a scenario that
is also found in the homogeneous Swift--Hohenberg
equation~\cite{Knobloch2005a,Avitabile2010d})

We also remark that, in a wide region of parameter space, $a \leq 1$, the kernel
is purely excitatory, yet snaking stable bumps are supported. 
When $a$ is further increased and the kernel becomes excitatory-inhibitory
($a>1$), the snaking limits become wider and involve solutions with multiple
threshold crossings.
We note that with a modulated but translation-invariant kernel, \revised{with modulation function} $A(x-y)$,
the integral over the resulting kernel, $W(x)$ would be monotonically increasing and
would then prevent the formation of stable bumps for $a<1$ \cite{Laing2003q}. The
inhomogeneity is thus a key ingredient to produce stable solutions in the absence of
inhibition when $a<1$.

The 
analytical methods presented in this paper could be useful in the future to study
time-periodic spatially-localized structures (often termed
\textit{oscillons}). A simple mechanism to obtain oscillatory instabilities in
neural field models is by introducing linear adaptation \cite{Pinto2001h}.
This modification seems amenable to study oscillons, since localized bumps of the
extended system can be constructed in the same way presented in this paper, yet the
corresponding stability problem changes slightly and may lead to a Hopf
bifurcation of the localized steady states. This approach has recently been used
by Folias and Ermentrout~\cite{Folias2012a} and Coombes and
co-workers~\cite{Coombes2013q} in two component models supporting breathers and
other spatio-temporal patterns.
Another possible extension is to study the effect of spatial modulation in planar neural
field models, in which case one could build upon the interface method developed in~\cite{Coombes2012c}
for homogeneous planar neural fields. 

\section*{Acknowledgements}
We are grateful to C\'{e}dric Beaume, Alan Champneys, Steve Coombes, David Lloyd
and Bj\"{o}rn Sandstede for valuable comments on a draft of the manuscript.  

 \appendix

 \section{Cell reduction for spatially-periodic states}
 \label{sec:apndcell}
 
 We aim to show that $2\pi\epsi$-periodic solutions $q$ to the
 integral model satisfy
  \[
    q(x) = \int_{-\pi\epsi}^{\pi\epsi} \tilde w(|x-y|) A(y) f(q(y)) \diff y,
    \qquad x \in [-\pi\epsi,\pi\epsi) 
  \]
  where
  \[
    \widetilde{w}(|x-y|) = \frac{1}{2} \e^{-|x-y|} +
    \frac{\e^{-2\pi\epsi}}{1-\e^{-2\pi\epsi}} \cosh \left( |x-y| \right),
  \]
  which correspond to Equations~\eqref{eq:2pireduced} and~\eqref{eq:modkernel}
  in the main text.

 For exponential kernels $w(|x-y|) = \exp(-|x-y|)/2$,
 Equation~\eqref{eq:steadyState} can be rewritten as
 \begin{eqnarray*}
   q(x) & = & \sum_{m=-\infty}^{-1}
   \int_{(2m-1)\pi\epsi}^{(2m+1)\pi\epsi} \frac{1}{2} \e^{-x+y} A(y) f(q(x-y)) \diff y\\
   			    && + \sum_{m=1}^{\infty}
			    \int_{(2m-1)\pi\epsi}^{(2m+1)\pi\epsi} \frac{1}{2} \e^{x-y} A(y)
			    f(q(x-y)) \diff y\\
			    && + \int_{-\pi\epsi}^{\pi\epsi} \frac{1}{2} \e^{-|x-y|} A(y)
			    f(q(x-y)) \diff y.
 \end{eqnarray*}
 By setting $\eta = y -2m\pi\epsi$,
 \begin{eqnarray*}
   \hspace{-0.7cm}q(x) & = & \sum_{m=-\infty}^{-1}
   \int_{-\pi\epsi}^{\pi\epsi} \frac{1}{2} \e^{-x+\eta} \e^{2m\pi\epsi} A(\eta)
   f(q(x-\eta)) \diff \eta\\
   			    && + \sum_{m=1}^{\infty} \int_{-\pi\epsi}^{\pi\epsi}
			     \frac{1}{2} \e^{x-\eta} \e^{-2m\pi\epsi} A(\eta) f(q(x-\eta)) \diff \eta\\
			    && + \int_{-\pi\epsi}^{\pi\epsi} \frac{1}{2} \e^{-|x-\eta|} A(\eta)
			    f(q(x-\eta)) \diff \eta.
 \end{eqnarray*}
 Here, we have made use of the fact that $q(x)$ and $A(x)$ are
 $2\pi\epsi$-periodic functions. Since
 \[
   \sum_{m=-\infty}^{-1} \e^{2m\pi\epsi}  = \sum_{m=1}^{\infty}
   \e^{-2m\pi\epsi} = \frac{\e^{-2\pi\epsi}}{1-\e^{-2\pi\epsi}}.
 \]
 we obtain the reduced formulation~\eqref{eq:2pireduced} with amended kernel~\eqref{eq:modkernel}.

  \section{Explicit solutions for cross-threshold solutions $q(x)$}
 \label{sec:apndexpl}
 
 
An explicit solution for equation \eqref{eq:2piloc} with Heaviside nonlinearity and
kernel \eqref{eq:kernel} is found by carrying out a direct integration, which
gives
 
\begin{equation}
 \qct(x) =
\begin{cases}
 \left(\e^x + \dfrac{2 \e^{-2\pi\epsi}}{1-\e^{-2\pi\epsi}} \cosh x \right) \Xi(L)  &\mbox{if} \quad -\pi\epsi < x < -L/2 \\
 1+ \dfrac{a \epsi^2}{\epsi^2 +1} \cos \dfrac{x}{\epsi} + 2 \cosh x \left( \dfrac{\e^{-2\pi\epsi}}{1-\e^{-2\pi\epsi}} \Xi(L) + \Upsilon(L)  \right) & \mbox{if} \quad -L/2 < x < L/2 \\
  \left(\e^{-x} + \dfrac{2 \e^{-2\pi\epsi}}{1-\e^{-2\pi\epsi}} \cosh x \right) \Xi(L)  &\mbox{if} \quad L/2< x < \pi\epsi 
 
\end{cases}.
\end{equation}
Here,
\begin{equation}
 \Xi(L) = \left( \sinh \frac{L}{2} \left( 1+ \frac{a \epsi^2}{\epsi^2+1} \cos \frac{L}{2 \epsi} \right) + \frac{a \epsi}{\epsi^2 +1} \cosh \frac{L}{2} \sin \frac{L}{2\epsi} \right),
\end{equation}
and
\begin{equation}
  \Upsilon(L) = \e^{-L/2} \left( 1+\frac{a \epsi^2}{\epsi^2 +1} \cos \frac{L}{2 \epsi} - \frac{a\epsi}{\epsi^2 +1} \sin \frac{L}{2\epsi} \right).
\end{equation}
The bifurcation equation is thus given by
\begin{equation}
  h = \left(\e^{-L/2} + \frac{2 \e^{-2\pi\epsi}}{1-\e^{-2\pi\epsi}} \cosh \frac{L}{2} \right)\times \Xi(L).
\end{equation}



\bibliographystyle{elsarticle-num}
\bibliography{da-hs-PhysD-Snakes}






\end{document}